\documentclass[aps,prl,showpacs,twocolumn,superscriptaddress]{revtex4-1}
\usepackage{graphicx,color,amsmath,amssymb}

\begin{document}

\title{Breaking of ensemble equivalence in networks}

\author{Tiziano Squartini}
\affiliation{Lorentz Institute for Theoretical Physics, University of Leiden (The Netherlands)}
\affiliation{Institute for Complex Systems, University of Rome ``La Sapienza'' (Italy)}
\author{Joey de Mol}
\affiliation{Lorentz Institute for Theoretical Physics, University of Leiden (The Netherlands)}
\affiliation{Mathematical Institute, University of Leiden (The Netherlands)}
\author{Frank den Hollander}
\affiliation{Mathematical Institute, University of Leiden (The Netherlands)}
\author{Diego Garlaschelli}
\affiliation{Lorentz Institute for Theoretical Physics, University of Leiden (The Netherlands)}

\date{\today}

\begin{abstract}
It is generally believed that, in the thermodynamic limit, the microcanonical 
description as a function of energy coincides with the canonical description 
as a function of temperature. However, various examples of systems for which 
the microcanonical and canonical ensembles are not equivalent have been 
identified. A complete theory of this intriguing phenomenon is still missing. 
Here we show that ensemble nonequivalence can manifest itself also in 
random graphs with topological constraints. We find that, while graphs with 
a given number of links are ensemble-equivalent, graphs with a given degree 
sequence are not. This result holds irrespective of whether the energy is 
nonadditive (as in unipartite graphs) or additive (as in bipartite graphs). In 
contrast with previous expectations, our results show that: (1) physically, 
nonequivalence can be induced by an extensive number of local constraints, 
and not necessarily by long-range interactions or nonadditivity; (2) mathematically, 
nonquivalence is determined by a different large-deviation behaviour of microcanonical 
and canonical probabilities for a single microstate, and not necessarily for almost all 
microstates. The latter criterion, which is entirely local, is not restricted to networks 
and holds in general.
\end{abstract}


\maketitle

\noindent
\emph{Background.}
In statistical physics, calculating the equilibrium properties of a system with a given
energy requires averaging over the so-called \emph{microcanonical ensemble}
\cite{Boltzmann1877,Gibbs1902}, i.e., the uniform distribution on the set of all 
particle configurations having a prescribed energy. Apart from trivial examples, this 
is a mathematically challenging task. Moreover, it is difficult to physically realize a 
situation where there is no uncertainty in the energy of the system. Therefore, it is 
often preferable to work with the so-called \emph{canonical ensemble} \cite{Gibbs1902}, 
i.e., a probability distribution with maximal entropy on an extended set of  configurations 
that `violate' the desired energy, but in such a way that the \emph{average} energy 
matches the prescribed value. This is achieved through an appropriate temperature, 
mathematically arising as the Lagrange multiplier enforcing the prescribed average 
energy. 

Starting with the work of Gibbs \cite{Gibbs1902}, the microcanonical and canonical 
ensembles have been shown to be equivalent in the thermodynamic limit (i.e., when 
the number of particles in the system tends to infinity) for physical systems with 
\emph{short-range interactions}. The original argument is that in the canonical ensemble 
at fixed temperature the energy fluctuations are negligible with respect to the average 
energy, so that in the thermodynamic limit the canonical ensemble is effectively 
microcanonical with a sharp value of the energy. Today, most textbooks in statistical 
physics still convey the message that equivalence of ensembles holds universally 
for \emph{every} physical system, justifying the use of energy and temperature as 
two different parameters giving an equivalent description. 

However, in the past decades various studies have highlighted that ensemble 
equivalence breaks down in certain models of fluid turbulence \cite{Ellis2000,Ellis2002}, 
quantum phase separation \cite{Blume1971,Barre2001,Ellis2004}, star formation 
\cite{LyndenBell1999,Chavanis2003}, nuclear fragmentation \cite{DAgostino2000}, 
and networks \cite{Barre2007,RS,myunbiased}. Physically, it is believed that nonequivalence is associated with \emph{long-range interactions} or other forms of 
nonadditivity \cite{ruffo}. However, a complete theoretical understanding of the
phenomenon is still missing. 
Mathematically, ensemble nonequivalence has been approached in various ways 
\cite{Touchette2004,touchette2014}. In particular, the microcanonical and canonical 
ensembles are said to be \emph{thermodynamically equivalent} \cite{Ellis2004} when 
the entropy and the free energy of the system are one-to-one related via a Legendre 
transform. The ensembles are said to be \emph{macrostate equivalent} \cite{Touchette2004} 
when the sets of equilibrium values of the macrostate (energy, magnetization, etc.) are 
the same. Finally, a recent and mathematically appealing definition is that of \emph{measure 
equivalence} \cite{touchette2014},  according to which the ensembles are said to be 
equivalent when the canonical probability distribution converges to the microcanonical 
probability distribution in the thermodynamic limit. Under certain \emph{hypotheses}, the 
three definitions have been shown to be equivalent \cite{touchette2014}. Moreover, large 
deviation theory \cite{LDT} shows that the ensembles are nonequivalent on all three levels 
when the microcanonical specific entropy is nonconcave as a function of the energy 
density in the thermodynamic limit \cite{touchette2014}. 

Here we study ensemble nonequivalence for \emph{networks with topological constraints} 
\cite{parknewman,gin3,squartini2011}. Usually, ensemble nonequivalence is studied for 
systems in which the Boltzmann distribution describes a certain physical interaction that 
is encapsulated in the energy. However, as already shown by Jaynes \cite{Jaynes}, the 
Boltzmann distribution describes much more general ensembles of systems with given 
constraints, namely, \emph{all solutions to the maximum-entropy problem of inference 
from partial information}. We argue that, for any discrete enumeration problem where 
we need to count microcanonical configurations compatible with a given constraint, there 
exists a `dual' problem involving canonical configurations induced by the same constraint. 
We prove a general result relating measure equivalence to equivalence of the large deviation 
properties of microcanonical and canonical probabilities, and provide examples of networks 
that exhibit nonequivalence whenever the number of constraints is extensive.

\noindent
\emph{Ensembles.}
For $N \in \mathbb{N}$, let $\mathcal{G}_N$ denote the discrete set of all configurations 
with $N$ particles (in the examples below, all graphs with $N$ nodes). Let $\vec{C}$ denote 
a vector-valued function on $\mathcal{G}_N$. The \emph{microcanonical distribution} with 
\emph{hard constraint} $\vec{C}^\star$ is defined as
\begin{equation}
P_{\mathrm{mic}}(\mathbf{G}) =
\left\{
\begin{array}{ll} 
1/\Omega_{\vec{C}^\star}, \quad & \text{if } \vec{C}(\mathbf{G}) = \vec{C}^\star, \\ 
0, & \text{else},
\end{array}\right.
\label{eq:PM}
\end{equation}
where $\Omega_{\vec{C}^\star} = | \{\mathbf{G} \in \mathcal{G}_N\colon\, \vec{C}(\mathbf{G}) 
= \vec{C}^\star \} |$ is the number of configurations that realize $\vec{C}^\star$. Following 
Jaynes \cite{Jaynes}, we introduce a `dual' problem involving a canonical probability 
distribution $P_{\mathrm{can}}(\mathbf{G})$ defined as the solution of the maximization 
of the Shannon entropy $S_N(P_{\mathrm{can}}) = - \sum_{\mathbf{G} \in \mathcal{G}_N} 
P_{\mathrm{can}}(\mathbf{G}) \ln P_{\mathrm{can}}(\mathbf{G})$ subject to the \emph{soft 
constraint} $\langle \vec{C} \rangle  = \vec{C}^\star$, where $\langle \cdot \rangle$ denotes the 
average w.r.t.\ $P_{\mathrm{can}}$, and to the normalization condition $\sum_{\mathbf{G} 
\in \mathcal{G}_N} P_{\mathrm{can}}(\mathbf{G})= 1$ \cite{parknewman}. This gives
\begin{equation}
P_{\mathrm{can}}(\mathbf{G}) = \frac{\exp[-H(\mathbf{G},\vec{\theta}^\star)]}{Z(\vec{\theta}^\star)},
\label{eq:PC}
\end{equation}
where $H(\mathbf{G}, \vec{\theta}) \equiv \vec{\theta} \cdot \vec{C}(\mathbf{G})$ is the 
\emph{Hamiltonian} and $Z(\vec{\theta}) \equiv \sum_{\mathbf{G} \in \mathcal{G}_N } 
\exp[-H(\mathbf{G}, \vec{\theta})]$ is the \emph{partition function}. Note that in eq.(\ref{eq:PC}) 
the parameter $\vec{\theta}$ must be set to the particular value $\vec{\theta}^\star$ that realizes 
$\langle \vec{C} \rangle  = \vec{C}^\star$ \cite{squartini2011}. This value also maximizes the 
likelihood \cite{mylikelihood}.

\noindent
\emph{Specific relative entropy and large deviations.}
The relative entropy of $P_{\mathrm{mic}}$ w.r.t.\ $P_{\mathrm{can}}$ is
\begin{equation}
S_N(P_{\mathrm{mic}} || P_{\mathrm{can}})  
= \sum_{\mathbf{G} \in \mathcal{G}_N} P_{\mathrm{mic}}(\mathbf{G}) 
\ln \frac{P_{\mathrm{mic}}(\mathbf{G})}{P_{\mathrm{can}}(\mathbf{G})}.
\label{eq:KL1}
\end{equation}
Following \cite{touchette2014}, we say that the two ensembles are measure equivalent 
if their \emph{specific relative entropy} is zero: 
\begin{equation}
s = \lim_{N \to \infty} \frac{S_N(P_{\mathrm{mic}} || P_{\mathrm{can}})}{N}=0.
\label{eq:criterion1}
\end{equation}

Before considering specific cases, we make a crucial observation. Noting 
from the form of $H(\mathbf{G},\vec{\theta})$ that  $P_{\mathrm{can}}(\mathbf{G}_1)
=P_{\mathrm{can}}(\mathbf{G}_2)$ when $\vec{C}(\mathbf{G}_1)=\vec{C}(\mathbf{G}_2)$ 
(the canonical probability is the same for all configurations with the same value 
of the constraint), we rewrite eq.(\ref{eq:KL1}) as
\begin{equation}
S_N(P_{\mathrm{mic}} || P_{\mathrm{can}}) 
= \ln \frac{P_{\mathrm{mic}}(\mathbf{G}^\star)}{P_{\mathrm{can}}(\mathbf{G}^\star)},
\label{eq:KL2}
\end{equation}
where $\mathbf{G}^\star$ is \emph{any}  configuration in $\mathcal{G}_N$ such that 
$\vec{C}(\mathbf{G}^\star) =\vec{C}^\star$. The equivalence condition in (\ref{eq:criterion1}) 
then becomes  
\begin{equation}
\lim_{N \to \infty} \frac{1}{N} \big[\ln {P_{\mathrm{mic}}(\mathbf{G}^\star)}
- \ln{P_{\mathrm{can}}(\mathbf{G}^\star)}\big] = 0,
\label{eq:criterion3}
\end{equation}
which demonstrates that nonequivalence coincides with $P_{\mathrm{mic}}(\mathbf{G}^\star)$ 
and $P_{\mathrm{can}}(\mathbf{G}^\star)$ having different large deviation behavior \cite{LDT}. 
Importantly, this  condition is purely local as it involves the microcanonical and canonical 
probabilities of a \emph{single} microstate $\mathbf{G}^\star$ realizing the hard constraint. 
This  greatly simplifies previously studied global conditions involving almost all microstates 
\cite{touchette2014}.

\noindent
\emph{Unipartite networks.}
We now apply the above concepts to the class of unipartite graphs, where there is a single 
set of nodes among which all possible links are allowed.

Let us first consider graphs with a fixed number of links $L$, i.e., $\vec{C} \equiv L$. Writing 
$L=\lambda V$, where $V\equiv N(N-1)/2$ is the number of pairs of nodes and $\lambda$ is 
the fraction of realized links, in the microcanonical ensemble we have
\begin{equation}
\Omega_{L^\star} = { V \choose L^\star }={ V \choose \lambda^\star V },
\qquad 0 < \lambda^\star < 1.
\label{eq:OmegaRG}
\end{equation}
The canonical ensemble can be obtained from eq.(\ref{eq:PC}) by setting $H(\mathbf{G},
\theta)=\theta L(\mathbf{G})$ and $p^\star \equiv \frac{e^{-\theta^\star}}{1 + e^{- \theta^\star}}
=\lambda^\star$ \cite{squartini2011}. This produces the \emph{Erd\H{os}-R\'enyi random 
graph} where each pair of nodes is connected with equal probability $p^\star$:
\begin{equation}
P_{\mathrm{can}}(\mathbf{G}) = (p^\star)^{L(\mathbf{G})} (1-p^\star)^{V - L(\mathbf{G})}.
\label{eq:PCRG}
\end{equation}
We can now compute the relative entropy from eq.(\ref{eq:KL2}) as
\begin{eqnarray}
S(P_{\mathrm{mic}} || P_{\mathrm{can}})
&=& 
- \lambda^\star V \ln \lambda^\star - (1 - \lambda^\star)V \ln (1 - \lambda^\star)\nonumber\\
- \ln { V \choose \lambda^\star V }
&=&\ln\sqrt{2\pi\lambda^\star(1-\lambda^\star)V}+O(1/V),
\end{eqnarray}
where we have used Stirling's formula $n! = (n/e)^n\sqrt{2\pi n}$ $[1+O(1/n)]$,
$n\to\infty$. This gives 
\begin{equation}
s = \lim_{N \to \infty} \frac{\ln\sqrt{2\pi\lambda^\star(1-\lambda^\star)V}}{N}=0,
\label{eq_s0}
\end{equation}
proving ensemble equivalence. In other words, when $N\to\infty$ most graphs have a 
number of links that is close to the average number of links.

We next consider graphs with a fixed degree sequence, i.e. $\vec{C} = \vec{k}
=(k_1, \dots, k_N)$ where $k_i$ is the number of links of node $i$. This is known 
as the \emph{configuration model} \cite{parknewman}. The microcanonical number 
$\Omega_{\vec{k}^\star}$ is not known in general, but asymptotic results exist in 
the `sparse case' where
\begin{equation}
k_{\mathrm{max}}  = \max_{1 \leq i \leq N} k_i = o( \sqrt{N}).
\label{eq:kmax}
\end{equation}
In this regime it is known that \cite{bender1977,mckay1991}
\begin{equation} 
\Omega_{\vec{k}^\star} = \frac{\sqrt{2}\,(\frac{2L^\star}{e})^{L^\star}}{\prod_{i=1}^N k^\star_i !}
\,e^{ - (\overline{k^{\star 2}}/2\overline{k^\star})^2 + \frac{1}{4} + o(N^{-1}\overline{k^\star}^{\,3}) }, 
\label{eq:omegaCM}
\end{equation}
where $\overline{k^\star} = N^{-1} \sum_{i=1}^N k^\star_i$ (average degree), $L=N\overline{k^\star}
/{2}$ (number of links), and $\overline{k^{\star 2}}= N^{-1} \sum_{i=1}^N k^{\star 2}_i$ (average square 
degree). The canonical ensemble is described \cite{squartini2011} by eq.(\ref{eq:PC}) where 
$H(\mathbf{G},\vec{\theta})=\vec{\theta}\cdot\vec{k}(\mathbf{G})$ and $\vec{\theta}^\star$ 
is such that 
\begin{equation}
\sum_{j\ne i} \frac{e^{-\theta^\star_i-\theta^\star_j}}{1+e^{-\theta^\star_i-\theta^\star_j}}=k^\star_i
\quad\forall\,i.
\label{eq:CM}
\end{equation}
Setting $p^\star_{ij}\equiv e^{-\theta^\star_i-\theta^\star_j}/(1+e^{-\theta^\star_i-\theta^\star_j})$, 
we have
\begin{equation}
P_{\mathrm{can}}(\mathbf{G}) 
= \prod_{i,j} (p^\star_{ij})^{g_{ij}} (1 - p^\star_{ij})^{1 - g_{ij} },
\label{eq:PCCM}
\end{equation}
where $\prod_{i,j}\equiv\prod_{i=1}^N\prod_{j<i}$ and $g_{ij}$ is the entry of the adjacency 
matrix of $\mathbf{G}$. Eq.(\ref{eq:kmax}) ensures that $k_{\mathrm{max}} = o(\sqrt{L})$, a 
condition under which eq.(\ref{eq:CM}) is solved as \cite{squartini2011}
\begin{equation}
p^\star_{ij} \sim e^{-\theta^\star_i-\theta^\star_j}=\frac{k^\star_i k^\star_j}{2L^\star} = o(1),
\label{eq:pijCL}
\end{equation}
where $\sim$ means that the quotient tends to $1$. This implies $\theta^\star_i \sim 
-\ln(k^\star_i/\sqrt{2L^\star})$ and $\ln(1-p^\star_{ij}) \sim - k^\star_i k^\star_j/2L^\star$. Thus
\begin{eqnarray}
\ln P_{\mathrm{can}}(\mathbf{G}^\star) 
&\sim& \sum_{i = 1}^N k^\star_i \ln k^\star_i - L^\star \ln (2L^\star) - L^\star. 
\label{eq:PCCL}
\end{eqnarray}
Combining eqs.(\ref{eq:KL2}), (\ref{eq:omegaCM}) and (\ref{eq:PCCL}), we obtain 
\begin{equation} 
\begin{aligned}
&S(P_{\mathrm{mic}} || P_{\mathrm{can}}) 
\sim  \sum_{i=1}^N \ln q(k^\star_i)\\ 
&\qquad + (\overline{k^{\star 2}}/2\overline{k^\star})^2
- \tfrac14 + o\big(N^{-1}\overline{k^\star}^{\,3}\big),
\end{aligned}
\end{equation}
where $q(k)\equiv k!/(k/e)^{k} \geq \sqrt{2\pi k}$ for $k\ge 1$. Eq.(\ref{eq:kmax}) guarantees that 
the terms in the last line are $o(N)$. Denoting a limiting average over nodes with a bar, we arrive at
\begin{equation}
s = \overline{ \ln q(k^\star)}
\geq \overline{\ln\sqrt{2\pi k^\star}} > 0,
\label{eq:limCL}
\end{equation}
proving nonequivalence. In other words, when $N\to\infty$ most 
graphs in the canonical ensemble do \emph{not} have a degree sequence that is close to the average degree sequence.
This important result explains various recent findings, e.g. the fact that the canonical and microcanonical entropies of random regular graphs 
are different even in the thermodynamic limit \cite{gin3} and that canonical fluctuations do not 
vanish in networks with local constraints \cite{myunbiased}. 

As a first example we consider \emph{sparse regular networks}, where every node has the same 
degree $k^\star=o(\sqrt{N})$. Then $\overline{\ln k^\star}=\ln k^\star$, so that eq.(\ref{eq:limCL}) 
becomes
\begin{equation}
s \geq \ln\sqrt{2\pi k^\star},\qquad k^\star=o(\sqrt{N}).
\label{eq:limRRG}
\end{equation}
Note that when $k^\star$ grows with $N$, $s$ diverges like 
$\ln k^\star$, signalling an extreme violation of equivalence. 

As a second example we consider \emph{sparse scale-free networks} \cite{cutoff}, defined by a 
truncated power-law degree distribution of the form $F_N(k) \equiv N^{-1} \sum_{i=1}^N 
1_{\{k_i = k\}} = A_\gamma k^{-\gamma}$ with $\gamma \in (1,\infty)$ for $1\leq k < k_c(N)$ and $F_N(k)=0$ for $k \geq k_c(N)$, 
where $\lim_{N\to\infty} k_c(N)=\infty$ 
and $k_c(N) = o(\sqrt{N})$. This `structural cut-off' \cite{cutoff} ensures eq.(\ref{eq:kmax}), so 
that eq.(\ref{eq:pijCL}) is valid. Approximating $F_N( k)$ by a continuous distribution, we 
see that the normalization of $F_N$ implies $A_\gamma\approx \gamma-1$, and so 
eq.(\ref{eq:limCL}) leads to
\begin{equation}
s \geq \overline{\ln\sqrt{2\pi k^\star}} \approx \frac{1}{2(\gamma-1)}+\ln\sqrt{2\pi},
\end{equation}
confirming nonequivalence. As the tail exponent $\gamma$ decreases, the degree distribution 
broadens and the degree of violation of equivalence increases.

Taken together, the above examples  indicate that ensemble equivalence holds 
when there is a single global constraint, while it is broken when there is an extensive number 
of local constraints. They also indicate that graphs with local constraints are always nonequivalent, irrespective of the breadth of the degree distribution.

\noindent
\emph{Bipartite networks.}
We now consider bipartite networks, where there are two distinct sets of nodes, and links are 
allowed only between the two sets. A bipartite graph $\mathbf{G}$ is specified by an $N\times M$ 
matrix, where $N$ and $M$ denote the numbers of nodes in the two sets. For simplicity, we 
constrain the topological properties on only one set (say, the one with $N$ nodes) and regard 
the other set as an `external environment'. Thus $N$ is the size of the system and the 
criterion in eq.(\ref{eq:criterion1}) still applies. 
For instance, we can think of our bipartite graph as a collaboration network of $N$ articles and $M$ authors. We may want to focus only on the properties of the set of articles, while regarding the `external' set of authors  fixed. In particular, in the limit $N\to\infty$ we may think of $M$ as a fixed number (either finite or infinite).

If we fix only the total number $L$ of links, then the number of microcanonical configurations is still 
given by eq.(\ref{eq:OmegaRG}) and the canonical probability, defined via $H(\mathbf{G},\theta)
=\theta L(\mathbf{G})$, is the same as in eq.(\ref{eq:PCRG}), where now $V=NM$ and 
$p^\star\equiv e^{-\theta^\star}$ is such that $\langle L\rangle=p^\star V=L^\star$. A calculation 
similar to that leading to eq.(\ref{eq_s0}) shows that $s=0$, proving again ensemble equivalence.

We next fix the degree sequence $\vec{k}^\star = (k_1^\star,\ldots,k_N^\star)$ of the constrained 
set. The microcanonical configurations are enumerated exactly as $\Omega_{\vec{k}^\star}=
\prod_{i=1}^N {M\choose k_i^\star}$. The canonical ensemble is defined by the Hamiltonian 
$H(\mathbf{G},\vec{\theta}) = \vec{\theta} \cdot \vec{k}(\mathbf{G})$ and is still described by 
eq.\eqref{eq:PCCM}, where $\prod_{i,j}=\prod_{i=1}^N\prod_{j=1}^M$ and $p^\star_{ij}
={k^\star_i }/{M}$ \cite{tiziano_bipartite}. We assume that $0<k^\star_i<M$ for all $i$ to avoid either disconnected nodes or fully connected nodes. 
A direct calculation yields
\begin{equation}
s = \overline{\ln\sqrt{ {2\pi k^\star(1-k^\star\!/M)}}}
\label{eq:bipartite}
\end{equation} 
(where the bar again denotes a limiting average, now over the $N$ nodes of the constrained set), 
proving ensemble nonequivalence. Note that here we have put no restriction on ${k}_\textrm{max}$, 
apart from requiring $0<{k}_\textrm{max}<M$. 
Indeed, while eq.\eqref{eq:limCL} is valid only in the sparse regime, eq.\eqref{eq:bipartite} holds in the full range of connectivity.

\noindent
\emph{Irrelevance of (non)additivity.} 
In the physics literature a connection has been conjectured between ensemble 
nonequivalence and the nonadditivity of the energy, as induced for instance by long-range 
interactions \cite{ruffo}. By contrast, we now show that in our examples 
the \emph{only mechanism} leading to nonequivalence is the presence of an \emph{extensive 
number of local constraints}, irrespective of (non)additivity. To this end, we partition the set of $N$ nodes into two sets $\mathcal{V}_1$ 
and $\mathcal{V}_2$ with $N_1=\alpha_1 N$ and $N_2=\alpha_2 N$ nodes, respectively, 
where $\alpha_1,\alpha_2>0$ and $\alpha_1+\alpha_2=1$. For a given graph $\mathbf{G}$, we calculate
the interaction energy between the two subsystems as $H_\textrm{int}(\mathbf{G},\vec{\theta})
= H(\mathbf{G},\vec{\theta})-H_1(\mathbf{G},\vec{\theta})-H_2(\mathbf{G},\vec{\theta})$, 
where $H_i(\mathbf{G},\vec{\theta})$ denotes the restriction of $H(\mathbf{G},\vec{\theta})$ 
to the set $\mathcal{V}_i$. 

In our example of unipartite graphs with the single constraint 
$\vec{C}=L$, the interaction energy is $H_\textrm{int}(\mathbf{G},\theta)=\theta[ L(\mathbf{G})
-L_1(\mathbf{G})-L_2(\mathbf{G})]$, where $L_i(\mathbf{G})$ is the number of `internal' links 
among the nodes of $\mathcal{V}_i$. Thus $H_\textrm{int}(\mathbf{G},\theta)$ is 
proportional to the number of links between $\mathcal{V}_1$ 
and $\mathcal{V}_2$, and its expected 
value is 
\begin{eqnarray}
\langle H_\textrm{int}(\theta)\rangle
&=&\theta \frac{p}{2}\big[{N(N-1)}-{N_1(N_1-1)}-{N_2(N_2-1)}\big]\nonumber\\
&=&\theta \frac{p}{2} N^2(1-\alpha_1^2-\alpha_2^2)=\theta p N^2\alpha_1\alpha_2
\end{eqnarray}
where $p = e^{-\theta}/(1+e^{-\theta})$ as in eq.(\ref{eq:PCRG}).
In the thermodynamic limit, the ratio of $\langle H_\textrm{int}(\theta)\rangle$ to the expected 
total energy $\langle H_\textrm{tot}(\theta)\rangle =\theta p N(N-1)/2$ is
\begin{equation}
\lim_{N\to\infty}\frac{\langle H_\textrm{int}(\theta)\rangle}
{\langle H_\textrm{tot}(\theta)\rangle}=2\alpha_1\alpha_2>0,
\end{equation}
which proves nonadditivity due to long-range interactions \cite{ruffo}. A similar result holds 
for the configuration model. So, unipartite networks 
are always nonadditive, irrespective of whether they exhibit nonequivalence.

By contrast, our examples of bipartite networks are always additive, irrespective of whether they exhibit (non)equivalence. This occurs because, when partitioning the set of $N$ nodes (e.g. articles in our previous example), both $\mathcal{V}_1$ and $\mathcal{V}_2$ remain connected only to the $M$ `external' nodes (e.g. authors), and not among themselves. Indeed, the Hamiltonian only couples the $N$ nodes to the external environment and we always get $H_\textrm{int}(\mathbf{G},\theta)=0$.  

\noindent
\emph{Conclusion.} 
We found that (non)equivalence is determined by an entirely local criterion involving the 
large-deviation behavior of microcanonical and canonical probabilities of a \emph{single} microstate 
rather than of \emph{almost all} microstates (as generally expected \cite{touchette2014}). 
This result is entirely general and is not restricted to networks.
Moreover, we found that the presence of an extensive number of local constraints provides a mechanism 
for ensemble nonequivalence in ensembles of graphs. While, in all examples known so far, nonequivalence 
was expected to be associated with long-range interactions or nonadditivity, in our examples (non)equivalence 
is only determined by the number of constraints, \emph{irrespective of (non)additivity}. 
From a practical point of view, graphs with local constraints are routinely used as null models to detect empirical patterns or to reconstruct networks from partial information \cite{myunbiased,parknewman,gin3,squartini2011,tiziano_bipartite}. 
So far, choosing between microcanonical and canonical implementations \cite{myunbiased} of these null models has been perceived as a mere matter of convenience. However, our findings imply that one should make a careful and principled choice, as results obtained using different ensembles may differ substantially. The same considerations might extend to other ensembles of systems with many constraints, applications of which range from biology (e.g. conformational ensembles) to finance and neuroscience (e.g. time series ensembles).

\begin{acknowledgments}
TS is supported by the Italian PNR project CRISIS-Lab. FdH is supported by ERC Advanced 
Grant 267356-VARIS and NWO Gravitation Grant 024.002.003-NETWORKS. DG is supported 
by the EU project MULTIPLEX (contract 317532) and the Dutch Econophysics Foundation 
(Stichting Econophysics, Leiden).
\end{acknowledgments}


\end{document}